\documentclass[a4paper]{spie}
\usepackage{graphicx,epsfig,epsf,color,hhline,import}
\usepackage{dcolumn}
\usepackage{bm}
\usepackage{tensor,pbox}
\usepackage{braket}
\usepackage{epstopdf}
\usepackage{amsmath,amsfonts}
\usepackage{amssymb}
\usepackage{transparent}
\usepackage{textpos}
\usepackage{subcaption}
\usepackage[a4paper,top=25.4mm,bottom=49.5mm,left=19mm,right=19.5mm
]{geometry}

\usepackage{changes}
\usepackage{cancel}
\usepackage{color}

\newcommand{\bea}{\begin{eqnarray}}
\newcommand{\ea}{\end{eqnarray}}
\newcommand{\eea}{\end{eqnarray}}

\newcommand{\sumint}[1]

\begin{document}
\newcommand{\rem}[1]{\textcolor{red}{\sout{#1}}}
\newcommand{\add}[1]{\textcolor{blue}{\uline{#1}}}
\newcommand{\ri}{ i}
\newcommand{\re}{ e}
\newcommand{\bx}{{\bm x}}
\newcommand{\bd}{{\bm d}}
\newcommand{\be}{{\bm e}}
\newcommand{\br}{{\bm r}}
\newcommand{\bk}{{\bm k}}
\newcommand{\bA}{{\bm A}}
\newcommand{\bD}{{\bm D}}
\newcommand{\bE}{{\bm E}}
\newcommand{\bB}{{\bm B}}
\newcommand{\bI}{{\bm I}}
\newcommand{\bH}{{\bm H}}
\newcommand{\bL}{{\bm L}}
\newcommand{\bR}{{\bm R}}
\newcommand{\bZero}{{\bm 0}}
\newcommand{\bM}{{\bm M}}
\newcommand{\bX}{{\bm X}}
\newcommand{\bn}{{\bm n}}
\newcommand{\bs}{{\bm s}}
\newcommand{\bv}{{\bm v}}
\newcommand{\tbs}{\tilde{\bm s}}
\newcommand{\rSi}{{\rm Si}}
\newcommand{\beps}{\mbox{\boldmath{$\epsilon$}}}
\newcommand{\bGamma}{\mbox{\boldmath{$\Gamma$}}}
\newcommand{\bxi}{\mbox{\boldmath{$\xi$}}}
\newcommand{\rg}{{\rm g}}
\newcommand{\tr}{{\rm tr}}
\newcommand{\xmax}{x_{\rm max}}
\newcommand{\xb}{\overline{x}}
\newcommand{\pb}{\overline{p}}
\newcommand{\ra}{{\rm a}}
\newcommand{\rx}{{\rm x}}
\newcommand{\rs}{{\rm s}}
\newcommand{\rP}{{\rm P}}
\newcommand{\up}{\uparrow}
\newcommand{\down}{\downarrow}
\newcommand{\hc}{H_{\rm cond}}
\newcommand{\kb}{k_{\rm B}}
\newcommand{\cI}{{\cal I}}
\newcommand{\tit}{\tilde{t}}
\newcommand{\cE}{{\cal E}}
\newcommand{\cC}{{\cal C}}
\newcommand{\Ubs}{U_{\rm BS}}
\newcommand{\sech}{{\rm sech}}
\newcommand{\qq}{{\bf ???}}
\newcommand*{\etal}{\textit{et al.}}
\def\vec#1{\bm{#1}}
\def\ket#1{|#1\rangle}
\def\bra#1{\langle#1|}
\def\keps{\bm{k}\boldsymbol{\varepsilon}}
\def\dm{\boldsymbol{\wp}}
\newcommand{\pscal}[2]{\ensuremath{ \langle \, #1 \, \vert  \, #2 \, \rangle}} 
\newcommand{\dens}[2]{\ensuremath{ \vert \, #1 \, \rangle \langle \, #2 \, \vert}} 
\newcommand{\moy}[1]{\ensuremath{ \langle \, #1 \, \rangle}} 
\newcommand{\moyvec}[2]{\ensuremath{ \langle #2 \, | \, #1 \, | \, #2 \rangle}} 
\newcommand{\com}[2]{\ensuremath{ \left[  #1 , #2  \right] }} 
\newcommand{\acom}[2]{\ensuremath{ \left\{  #1 , #2  \right\} }} 
\newcommand{\paran}[1]{\ensuremath{ \left(  #1  \right) }} 
\newcommand{\jmoins}{$J_{-} \,$}
\newcommand{\jplus}{$J_{+} \,$}
\newcommand{\jx}{$J_{x} \,$}
\newcommand{\jy}{$J_{y} \,$}
\newcommand{\jz}{$J_{z} \,$}
\newcommand{\lu}{\textcolor{blue}}
\newcommand{\lug}{\textcolor{green}}
\renewcommand{\figurename}{Figure}


\title{A quantum-chaotic cesium-vapor magnetometer}
\author[a]{Lukas J.~Fiderer}
\author[a]{Daniel Braun}
\affil[a]{Institute for Theoretical Physics, University of T{\"u}bingen, Auf der Morgenstelle 14, 72076 T{\"u}bingen, Germany}

\authorinfo{Send correspondence to L.J.F. or D.B.\\ L.J.F.: E-mail: lukas.fiderer@uni-tuebingen.de\\  D.B.: E-mail: daniel.braun@uni-tuebingen.de}

\maketitle
\begin{abstract}
Quantum-enhanced measurements represent the path towards the best measurement precision allowed by the laws of quantum mechanics. Known protocols usually rely on the preparation of entangled states and promise high or even optimal precision, but fall short in real-word applications because of the difficulty to generate entangled states and to protect them against decoherence. Here, we refrain from the preparation of entangled states but supplement the integrable parameter-encoding dynamics by non-linear kicks driving the system in the dynamical regime of quantum chaos. We show  that large improvements in measurement precision are possible by modeling a spin-exchange relaxation-free alkali-vapor magnetometer where the non-linear kicks are realized by exploiting the ac Stark effect.  
\end{abstract}

\section{INTRODUCTION}
High-precision sensors are based more and more often on quantum systems such as NV centers\cite{mamin2013nanoscale}, photons in an interferometer\cite{lee2012interferometric}, or clouds of atoms\cite{peters2001high}. Optimizing such sensors allows measurement precision to reach the standard quantum limit. To overcome this limit, it is necessary to exploit genuine quantum properties such as entanglement of the quantum systems under consideration.  Such quantum-enhanced measurements were proposed to improve frequency standards
\cite{Huelga97,PhysRevLett.86.5870,Leibfried04,wasilewski2010quantum,koschorreck2010sub},       
navigation 
\cite{Giovannetti01}, remote sensing \cite{Allen08radar}, measurement
of magnetic 
fields \cite{Taylor08}, or gravitational wave detection 
\cite{Goda08,aasi2013enhanced}.

Common features of many propositions are the preparation of non-classical, entangled states and integrable dynamics of the quantum sensor\cite{pezze2018quantum}. Although the preparation of  entangled states is a theoretically well-established resource that in principle allows one to reach optimal measurement precision, in practice, it represents a huge challenge to generate entanglement over large systems that is necessary to achieve large improvements in measurement precisions. Further, entangled states are prone to decoherence which leads, for instance, to a severe decrease of squeezing in interferometers that operate with squeezed vacuum. In particular, it has been shown that in most cases decoherence prevents reaching the desirable Heisenberg scaling of precision such that the quantum improvement reduces to a constant factor over the standard quantum limit\cite{escher2011general,demkowicz2012elusive}.

Decoherence in quantum systems and experimental difficulties in generating entanglement call for quantum-enhanced protocols that perform well under decoherence without requiring the preparation of large-scale entanglement. The idea, we proposed in our work about quantum-chaotic sensors\cite{fiderer2018quantum}, is the following: Prepare the quantum sensor in a coherent state, i.e., the most classical state, that is typically easy to prepare. Then, the dynamics that encodes the parameter to be measured is supplemented with non-linear kicks, corresponding to some non-linear Hamiltonian that does not commute with the Hamiltonian of the parameter-encoding dynamics. This can create quantum chaos, i.e., dynamics that becomes chaotic in the classical limit. The unpredictability of classical chaos originating from its exponential sensitivity to a change of initial conditions is absent in the quantum case; instead quantum chaos is characterized by its sensitivity to a change of the dynamics which can be quantified by the Loschmidt echo. The Loschmidt echo, however, is in the limit of small changes of the dynamics related to the quantum Fisher information which quantifies the achievable measurement precision (a more precise description is given in section \ref{sec:qfi}), hinting at an underlying connection of quantum chaos and achievable measurement precision. At the example of the kicked top model, this correspondence has been studied numerically  in the decoherence-free regime as well as under decoherence\cite{fiderer2018quantum}.

Here we simulate a spin-exchange relaxation-free cesium-vapor magnetometer that is turned into a quantum-chaotic sensor by periodically applying nonlinear, short kicks to the cesium spins that precess in the magnetic field which has to be measured. Comparing the measurement performance with the same sensor in the absence of kicks, we find an improvement in measurement precision.

In section \ref{sec:kick}, we introduce the kicked top model, a standard model to study quantum chaos. In section \ref{sec:serf}, it is discussed how to turn an alkali-vapor magnetometer into a quantum-chaotic sensor based on the  kicked top model. The master equation for such a quantum-chaotic magnetometer is given in section \ref{sec:master}. Finally, in section \ref{sec:measurement}, we introduce the framework of quantum Fisher information which is then used to quantify and compare the measurement precision of the quantum-chaotic magnetometer.

\section{THE KICKED TOP MODEL}\label{sec:kick}
Consider an atomic spin of size $f$ with spin operator $\textbf{F}=(F_x,F_y,F_z)$, $F_z\ket{fm}=\hbar m\ket{fm}$ and $\textbf{F}^2\ket{fm}=\hbar^2 f(f+1)\ket{fm}$, where $f$ and $m$ are atomic angular momentum quantum numbers. For such an atomic spin, or actually for any quantum system fulfilling the angular momentum algebra, we can set up the kicked top model\cite{haake1987classical,kickedtop,haake2013quantum} with time-dependent Hamiltonian
\begin{equation}\label{eq:hamilton}
{H}_\text{KT}(t)=\alpha F_y+\frac{k}{(2f+1)\hbar}F_x^2\sum_{n=-\infty}^\infty \tau\delta(t-n\tau)\,.
\end{equation}
Setting the period $\tau=1$, the Floquet operator
\begin{align}
U_\alpha(k)=e^{-ik\frac{ F^2_x
	}{(2f+1)\hbar^2}}e^{-i\alpha \frac{F_y}{\hbar}}
\end{align}
describes the unitary dynamics from time $t$ to $t+\tau$.
The first term $\alpha F_y$ of the Hamiltonian generates a rotation or precession of the angular momentum around the $y$-axis by an angle $\alpha$. The second term contains the non-linearity $F_x^2$ generating an instantaneous kick only if time $t$ equals $n\tau$ where $n$ can be any integer. $k$ is the kicking strength. In total, the dynamics consist of continuous spin precession periodically disrupted by instantaneous nonlinear kicks. Figure \ref{fig:term} (c) shows the kicked top dynamics integrated in a measurement protocol measuring the precession angle $\alpha$.

The kicked top is a standard model in quantum chaos because its dynamics is non-integrable for $k\geq 0$ and its classical limit shows a transition from regular to chaotic dynamics when $k$ is increased. The kicked top has been realized with single-atom spins in a cold gas\cite{chaudhury2009quantum} as well as with a pair of spin-$1/2$ nuclei using NMR techniques\cite{krithika2018nmr}.

\section{A kicked cesium-vapor magnetometer}\label{sec:serf}
 Alkali-vapor magnetometers in the spin-exchange relaxation-free (SERF) regime\cite{allred2002high,Kominis03,savukov2005effects,budker2007optical,sheng2013subfemtotesla} typically consist of a glass cell filled with the alkali vapor with a density of about $10^{13}$ atoms per cm$^3$. The optically polarized spins of the alkali atoms are the quantum system of the magnetometer that reacts to a magnetic field with precession around the axis of orientation of the magnetic field. The dominant damping mechanisms originate from collisions of alkali atoms among each other and with the walls of the glass cell. Therefore, buffer gas is added to the cell and the inside walls are coated with an anti-relaxation coating. The SERF regime is determined by a spin-exchange rate that is much higher than the Larmor precession of the spins induced by the magnetic field to be measured; it is realized by small magnetic fields ($10^{-13}\,$T), high gas pressure in the cell, and heating of the atomic vapor.
 \subsection{Finding a proper parameter regime}
 In the following, we show how to implement the kicked top model in an alkali-vapor magnetometer in the spin-exchange relaxation-free (SERF) regime\cite{allred2002high,Kominis03,savukov2005effects,budker2007optical,sheng2013subfemtotesla}.
 We model a cesium-vapor magnetometer at room temperature
similar to the experiments with rubidium vapor of Balabas et al.~\cite{balabas2010polarized}. 

\begin{figure}
	\centering
	\begin{subfigure}{.5\textwidth}
		\centering
		\includegraphics[width=0.7\linewidth]{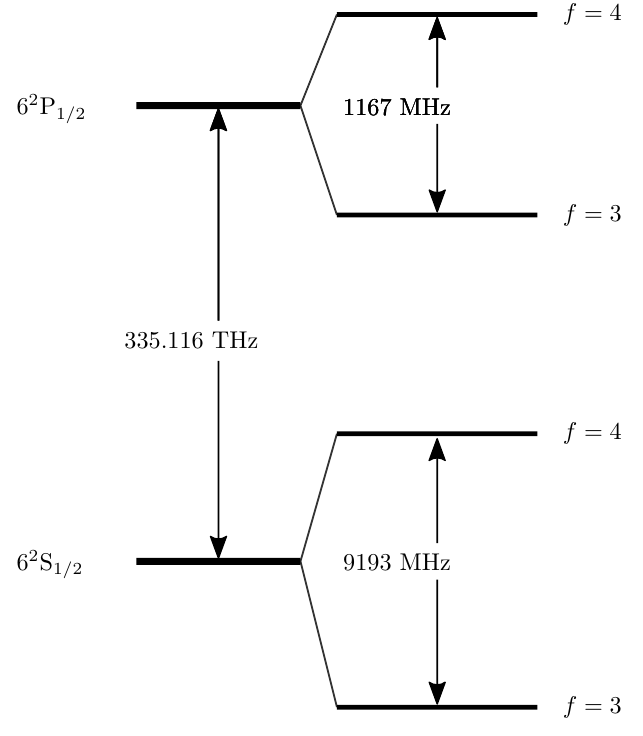}
		\label{fig:sub1}
	\end{subfigure}%
	\begin{subfigure}{.5\textwidth}
		\centering
		\includegraphics[width=\linewidth]{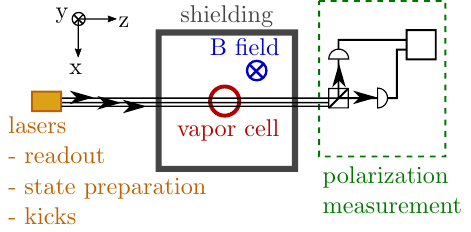}\\
		(b)\\\vspace*{1cm}
		
		\label{fig:sub2}
		\includegraphics[width=\linewidth]{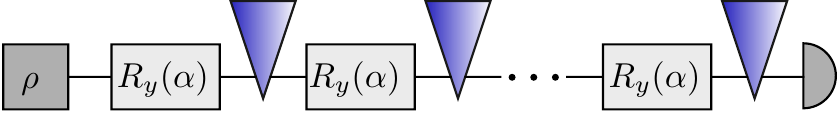}
	\end{subfigure}
\vspace*{0.3cm}\\
(a)\hspace{8cm}(c)\vspace*{0.5cm}
	\caption{Panel (a): Hyperfine structure of the D1 line of  $^{133}$Cs (not to scale). Panel (b): Schematic setup of a kicked cesium-vapor magnetometer. State preparation consists of polarizing the atomic spins optically; non-linear kicks and readout are realized optically as well, where a polarization measurement is performed for the transmitted readout beam. Panel (c): Schematic procedure of a quantum chaotic measurement. A state $\rho$ is prepared, precession $R_y(\alpha)$ encodes the parameter $\alpha$ to the system, non-linear kicks (blue wedges) are applied periodically, and final readout (semicircle) allows to estimate $\alpha$.}\label{fig:term}
\end{figure}

The Cs spin is composed of a nuclear spin $K=7/2$ and an electronic spin $s=1/2$ of a single valence electron
which splits the ground state
$6^2\text{S}_{1/2}$ into two energy levels with total spin $f_1=3$ and
$f_2=4$. This results in an effective Hilbert space of dimension
$2(2K+1)=16$ for our model of a kicked SERF magnetometer. Consider the D1 line of $^{133}$Cs as depicted in figure \ref{fig:term} (a). 
The kicks will be realized exploiting the ac Stark effect, a rank-2 light shift generated by linearly-polarized, off-resonant light pulse. More precisely, with respect to the $6^2$S$_{1/2}$, $f_1=3$ ground state, the light pulse is tuned halfway between the two hyperfine components of $6^2$P$_{1/2}$ because the light shift is strongest there\cite{deutsch2010quantum,fernholz2008spin}. The ac Stark effect is present only if the hyperfine splitting of $6^2$P$_{1/2}$ is resolved which is \textit{not} the case under conditions typical for SERF magnetometers due to pressure broadening. Therefore, we choose a density of only $2\times 10^{10}$ cesium atoms per cm$^3$ and no buffer gas. It was shown by Balabas et al.\cite{balabas2010polarized} that thanks to an alkene-based anti-relaxation coating of the walls of the glass cell it is still possible to reach the SERF regime. Then, pressure broadening is negligible compared to $357$\,MHz (FWHM) of Doppler broadening of the hyperfine levels of $6^2$P$_{1/2}$, which is  obtained from the temperature ($294$\,K, room temperature) and the mass of a cesium atom. Then, the relatively large hyperfine splitting of cesium  ($1167$\,MHz, figure \ref{fig:term} (a)) is large enough to be well resolved. Note that heating of the glass cell is counterproductive as it increases Doppler broadening.

This represents a parameter regime where magnetic precession of the cesium spins corresponds to the linear precession Hamiltonian in the kicked top model, and short off-resonant light pulses induce the ac Stark effect corresponding to the non-linear kicks. In the next section we give a master equation taking into account all relevant decoherence mechanisms including decoherence due to the off-resonant light pulses as well as the finite length of the light pulses in contrast to the idealized delta-shaped kicks of the kicked top model.
\subsection{Master equation}\label{sec:master}
 Collisions of cesium atoms lead to spin-exchange and spin-destruction processes with rates $R_\textrm{se}\simeq 12\,$Hz and $R_\textrm{sd}\simeq0.12\,$Hz, respectively. These decay rates can be estimated from the cross sections of Cs-Cs
collisions\cite{bhaskar1980spin}, the mean relative thermal velocity of cesium atoms and their density\cite{ledbetter2008spin}. The Larmor frequency is given by $\Omega_\text{Lar}=g_f\mu_\text{B}B/\hbar$ with the Land\'e g-factor $g_f$ and the Bohr magneton $\mu_\text{B}$, and we set the magnetic field strength to $B=4\times 10^{-14}$. In the electronic ground state $6^2$S$_{1/2}$, we have $\Omega_\text{Lar}\simeq 0.44$\,mHz such that $R_\textrm{se}\gg \Omega_\text{Lar}$, i.e., we are in the SERF regime. Then, the effective, joint dynamics of both hyperfine ground states of $6^2$S$_{1/2}$ is unaffected by spin-exchange collisions. The alkene-based anti-relaxation coating from Balabas et al.\cite{balabas2010polarized} supports up to $10^6$ collisions before the atomic spins become depolarized. For a spherical vapor-cell with $3$\,cm diameter, this leads to a damping rate due to collisions with the wall of $R_\textrm{wall}\simeq 11\,$mHz. Since $R_\textrm{sd}\gg R_\textrm{wall}$, spin destruction is the dominant damping mechanism, and we neglect damping due to collisions with the wall.

In order to model non-integrable dynamics, the theoretical description must take into account the full Hilbert space of the electronic ground state in contrast to the standard modeling of SERF magnetometers where the nuclear-spin component is eliminated. We use the following notation: the atomic spin $\textbf{F}=\textbf{K}+\textbf{J}$ can be decomposed in nuclear spin $\textbf{K}$ and electronic angular momentum $\textbf{J}=\textbf{L}+\textbf{S}$, where $\textbf{L}$ is the orbital angular momentum and $\textbf{S}$ the electron spin.   The evolution of the spin density matrix $\rho$ is described by a master 
equation that includes damping originating from collisions of cesium 
atoms\cite{appelt1998theory} and an interaction with an off-resonant 
light field in the low-saturation limit\cite{deutsch2010quantum} 
modeling the non-linear kicks:
\begin{align}\label{eq:SERFmaster}
\frac{d\rho}{dt}&=R_\textrm{se}\left[\varphi(1+4\braket{\textbf{S}}\cdot\textbf{S})-\rho\right]+R_{\text{sd}}\left[\varphi-\rho\right] 
+a_\textrm{hf}\frac{[\textbf{K}\cdot\textbf{S},\rho]}{i\hbar}+\frac{H_\text{A}^{\text{eff}}\rho-\rho H_\text{A}^{\text{eff}\dagger}}{i\hbar} 
\notag\\&\quad +\gamma_\text{nat}\sum_{q=-1}^1\left(\sum_{f,f_1}W_q^{ff_1}\rho_{f_1f_1}\left(W_q^{ff_1}\right)^\dagger+\sum_{f_1\neq f_2}W_q^{f_2f_2}\rho_{f_2f_1}\left(W_q^{f_1f_1}\right)^\dagger\right)\,.
\end{align}
The first summand describes spin-exchange 
relaxation, and $\varphi=\rho/4+\textbf{S}\cdot\rho\textbf{S}$ is called the purely 
nuclear part of the density matrix, where the electron-spin operator $\textbf{S}$ 
only acts on the electron-spin component with expectation value  
$\braket{\textbf{S}}=\tr[\textbf{S}\rho]$. Spin-destruction relaxation is given by the second summand.
The third summand describes the hyperfine coupling of nuclear spin \textbf{K} 
and electronic spin \textbf{S} with hyperfine structure constant $a_\text{hf}$, 
and the fourth summand drives the dynamics with an effective
non-hermitian Hamiltonian on both ground-state hyperfine manifolds $H_\text{A}^\text{eff}=H_{\text{A},f=3}^\text{eff}+H_{\text{A},f=4}^\text{eff}$, with
\begin{equation}
H_{\text{A},f}^\text{eff}=\hbar\Omega_\text{Lar}F_y+\sum_{f'} \frac{\hbar\Omega^2 C_{j'f'f}^{(2)}}{4(\Delta_{ff'}+i\gamma_\text{nat}/2)}\left(\boldsymbol{\epsilon_\text{L}}\cdot\textbf{F}\right)^2\,.
\end{equation}
The effective Hamiltonian includes Larmor precession with frequency $\Omega_\text{Lar}$ 
of the 
atomic spin in the external magnetic field $\textbf{B}=B\hat{\textbf{y}}$ and the rank-2 light-shift induced by a linearly 
polarized light pulse  with unit polarization vector $\boldsymbol{\epsilon}_\text{L}$ 
of the light field and off-resonant with detuning $\Delta_{ff'}$ from 
the D1-line transition with $f\rightarrow f'$. We take kick pulses to be polarized in $x$-direction, $\boldsymbol{\epsilon}_\text{L}=\hat{\textbf{x}}$, while the pulse beam propagates in $z$-direction, as indicated in figure \ref{fig:term} (b). 
Further, the prefactor of the non-linearity consists of the  characteristic Rabi frequency
$\Omega=\gamma_\text{nat}\sqrt{I_\text{kick}/(2I_\text{sat}) }$ of the D1 line, with saturation intensity of off-resonant linearly polarized light\cite{steck2003cesium} $I_\text{sat}\simeq 2.5$\,mW\,cm$^{-2}$, natural line width $\gamma_\text{nat}$, kick-laser intensity $I_\text{kick}$, and coefficients 
\begin{equation}
C_{j'f'f}^{(2)}=(-1)^{3f-f'}\frac{\sqrt{30}(2f'+1)}{\sqrt{f(f+1)(2f+1)(2f-1)(2f+3)}} \left\{\begin{matrix}
f & 1 & f'\\ 1 & f & 2 \end{matrix}\right\}\left|o_{1/2 f}^{j'f'}\right|^2.
\end{equation}
The curly braces denote the Wigner $6j$ symbol and $o_{jf}^{j'f'}$ is defined as
\begin{equation}\label{eq:ofunc}
o_{jf}^{j'f'}=(-1)^{f'+1+j'+K}\sqrt{(2j'+1)(2f+1)}\left\{\begin{matrix}
f & K & j'\\ j & 1 & f \end{matrix}\right\},
\end{equation}
where the total angular momenta of ground and excited levels of the D1 
line are $j=j'=1/2$.
Photon scattering is taken into account by the imaginary shift $i\gamma_\text{nat}/2$ of 
the detuning $\Delta_{ff'}$ in the effective Hamiltonian and by the remaining parts 
of the master equation that correspond to optical pumping which leads 
to cycles of excitation to the 6P$_{1/2}$ manifold followed by spontaneous 
emission to the ground-electronic manifold 6S$_{1/2}$. When the laser 
is switched off, the master equation solely involves the first four 
summands (spin exchange, spin destruction, hyperfine coupling, effective Hamiltonian) and the effective Hamiltonian reduces to the Larmor precession term.

The jump operators are given as\cite{deutsch2010quantum}
\begin{align}
W_q^{f_bf_a}=\sum_{f'=3}^4\frac{\Omega/2}{\Delta_{f_af'}+i\gamma_\text{nat}/2}\left(\textbf{e}_q^*\cdot\textbf{D}_{f_bf'}\right)\left(\boldsymbol{\epsilon}_\textrm{L}\cdot\textbf{D}_{f_af'}^\dagger\right),
\end{align}
with the spherical basis $\textbf{e}_1=-(\hat{\textbf{x}}+i\hat{\textbf{y}})/\sqrt{2}$,
$\textbf{e}_0=\hat{\textbf{z}}$,  $\textbf e_{-1}=(\hat{\textbf{x}}-i\hat{\textbf{y}})/\sqrt{2}$, where  $\hat{\textbf{x}}, \hat{\textbf{y}}, \hat{\textbf{z}}$ denotes the  Cartesian basis, and $\textbf{D}^\dagger_{ff'}=\sum_{q,m,m'}\textbf{e}_q^*o_{jf}^{j'f'}\braket{f'm'|fm;1q}\ket{f'm'}\bra{fm}$ is the raising operator with Clebsch-Gordan coefficients $\braket{f'm'|fm;1q}$.

The Euler method is used to solve the non-linear trace-preserving master equation and 
hyperfine coupling is taken into account
by setting off-diagonal blocks of the density matrix in the coupled 
$\ket{fm}$-basis after each Euler step to zero\cite{savukov2005effects}, because they oscillate 
very quickly with frequency $a_\text{hf}$. In distinction from the kicked top model, kicks are not 
assumed to be arbitrarily short, i.e., kicks and precession coexist
during a light pulse. In each Euler step during a kick, the non-linearity and the corresponding dissipation is factored in
by calculating a superoperator and applying it to the state.

The period between pulses, and the duration and intensity of light pulses  must be chosen such that the favorable effect of kicking outweighs the detrimental effect of induced decoherence. The intensity of the light pulses inducing the non-linear kicks is set to
$I_\text{kick}=0.1\,\text{mW/cm}^2$ and light pulses are detuned halfway between 
the hyperfine splitting of $6^2$P$_{1/2}$, $\Delta_{34}\simeq
-584\,\textrm{MHz}$.   
The period is $\tau=1\,\textrm{ms}$ where during the last 
$2\,\mu$s of each period the light pulse is applied which corresponds to a very low effective kicking strength of $k\simeq6.5\times 10^{-4}$ for the lower hyperfine level of the ground state.

Doppler broadening is taken into account by numerically averaging the righthand side of the master equation over the Maxwell-Boltzmann distribution of velocities of an cesium atom. This translates into an average over detunings $\Delta_{ff'}$. We limit averaging over detunings to a $3\sigma$ interval because $\Delta\gg\Omega$ must hold within the description of the master equation\cite{deutsch2010quantum}.

The initial spin-state is polarized with a circularly polarized pump beam in $z$-direction orthogonal to the magnetic field (see figure \ref{fig:term} (b)) resonant 
with the D1 line which in 
the presence of spin-relaxation leads to an effective thermal state\cite{appelt1998theory}
\begin{equation}
\rho=\frac{e^{\beta K_z}e^{\beta S_z}}{Z_K Z_S},
\end{equation}
where $\beta=\ln\frac{1+q}{1-q}$, 
with polarization $q=0.95$, and $Z_j=\sum_{m=-j}^je^{\beta m}$ is the partition function. The readout is accomplished typically  
with the help of an off-resonant probe beam by measuring its 
polarization which experienced a Faraday rotation during the interaction 
with the atomic spin ensemble, see figure \ref{fig:term} (b). Since we are interested only in a comparison of measurement strategies, with kicks and without kicks, we do not model readout noise which is the same for both strategies.
\section{MEASUREMENT PRECISION}\label{sec:measurement}
We will first introduce Fisher information and quantum Fisher information and show the connection of the latter to fidelity establishing a fundamental relation between achievable measurement precision and quantum chaos. Then, we will compare measurement strategies (kicks versus no kicks) by means of these quantities.
\subsection{Quantum Fisher information and fidelity}\label{sec:qfi}
If the probability distribution of measurement outcomes depends on a parameter $\alpha$, we can estimate $\alpha$ from the measurement outcomes. High measurement precision corresponds to a low variance (or standard deviation) of our estimate $\alpha_\text{est}$ of $\alpha$ which is lower bounded by the Cram\'er--Rao bound,
\begin{align}
\text{Var}(\alpha_\text{est}) \geq \frac{1}{M
	I_{\textrm{Fisher},\alpha}},
\end{align}
where $M$ denotes the number of measurements and $I_{\text{Fisher},\alpha}$ is the Fisher information\cite{Helstrom1969}. It is given by
\begin{align}\label{eq:Fisher}
I_{\text{Fisher},\alpha}:=\int d\xi\frac{(dp_\alpha(\xi)/d\alpha)^2}{p_\alpha(\xi)}\,,
\end{align}
where $p_\alpha(\xi)$ denotes the probability of observing the measurement outcome $\xi$. Describing the parameter-dependent quantum state of the quantum sensor by a density operator $\rho_\alpha$ and the measurement by a positive operator-valued measure (POVM) $\{\Pi_\xi\}$ with positive operators
$\Pi_\xi$ that correspond to observing the measurement outcome $\xi$, the probability for observing $\xi$ is given by the Born rule, $p_\alpha(\xi)=\tr\left[\Pi_\xi\rho_\alpha\right]$. Minimizing $\text{Var}(\alpha_\text{est})$ with respect to the choice of measurement, yields the quantum Cram\'er--Rao bound,
\begin{align}\label{eq:qfi_inegality}
\text{Var}(\alpha_\text{est}) \geq \frac{1}{M I_\alpha},
\end{align}
where $I_\alpha$ is the quantum Fisher information\cite{Braunstein94} (QFI),
\begin{align}
\label{eq:QFI}
I_\alpha=\lim_{\epsilon\to 0}4\frac{1-F_\epsilon}{\epsilon^2}\,,
\end{align}
with $F_\epsilon$ the fidelity between the state $\rho_\alpha$ and the perturbed state $\rho_{\alpha+\epsilon}$,
\begin{align}\label{eq:fidelity}
F_\epsilon=\left\lVert \sqrt{\rho_\alpha}\sqrt{\rho_{\alpha+\epsilon}}\right\rVert_2^2,
\end{align}
where $\lVert A \rVert_1=\tr\sqrt{AA^\dagger}$ is the trace norm.

In the field of quantum chaos, fidelity, usually evaluated for pure states and unitary dynamics, is called Loschmidt echo and represents an important quantity to study the sensitivity to changes of a parameter of the dynamics \cite{Gorin06review}. As can be seen from equation (\ref{eq:QFI}), only in the limit of small perturbations, known as the perturbative regime \cite{PhysRevE.64.055203,PhysRevE.65.066205}, QFI is given by the fidelity.

Comparing two measurement strategies by comparing the maximal QFI or Fisher information of both strategies is not fair in terms of resources if the maxima are reached at different times, i.e., if the two strategies consume different amounts of time. In principle, the faster measurement strategy could be followed up with another measurement until the other strategy is finished, or it might be better to not even wait until QFI is maximal but instead perform a series of shorter measurements\footnote{The underlying assumption is that it is always possible to interrupt and repeat a measurement and that preparation and readout times of single measurements can be neglected.}. This means that there is a tradeoff between measurement time and repetitions of a measurement which gives a prefactor $M$ for the QFI (equation \ref{eq:qfi_inegality}) for $M$ repetitions. Therefore, we consider the rescaled QFI and rescaled Fisher information,
 \begin{align}
I_\alpha^{(t)}&=
\frac{I_\alpha}{t},\\
I_{\textrm{Fisher},\alpha}^{(t)}&=
\frac{I_{\textrm{Fisher},\alpha}}{t},
 \end{align}
 with measurement time $t$. 
Rescaled QFI (or rescaled Fisher information) decays if QFI (Fisher information) is $\propto t^x$ with $x<1$ indicating that it is better to stop the measurement and start a new one; linear $M$ scaling is then better than $\propto t^x$ scaling. Classical averaging over long times typically leads to a $x=1$ scaling which motivates giving sensitivity (standard deviation of the estimator) in units of $1/\sqrt{\text{Hz}}$ which is well established in experimental physics. On the other hand, quantum coherence typically leads to an $x=2$ scaling.  
Rescaled QFI allows us to compare measurement strategies that use different amounts of time; best precision corresponds to the maximum 
rescaled QFI or Fisher information.
\subsection{Comparison of a cesium-vapor magnetometer with and without kicks}
\begin{figure}[h!]
		\centering
		\includegraphics[width=9cm]{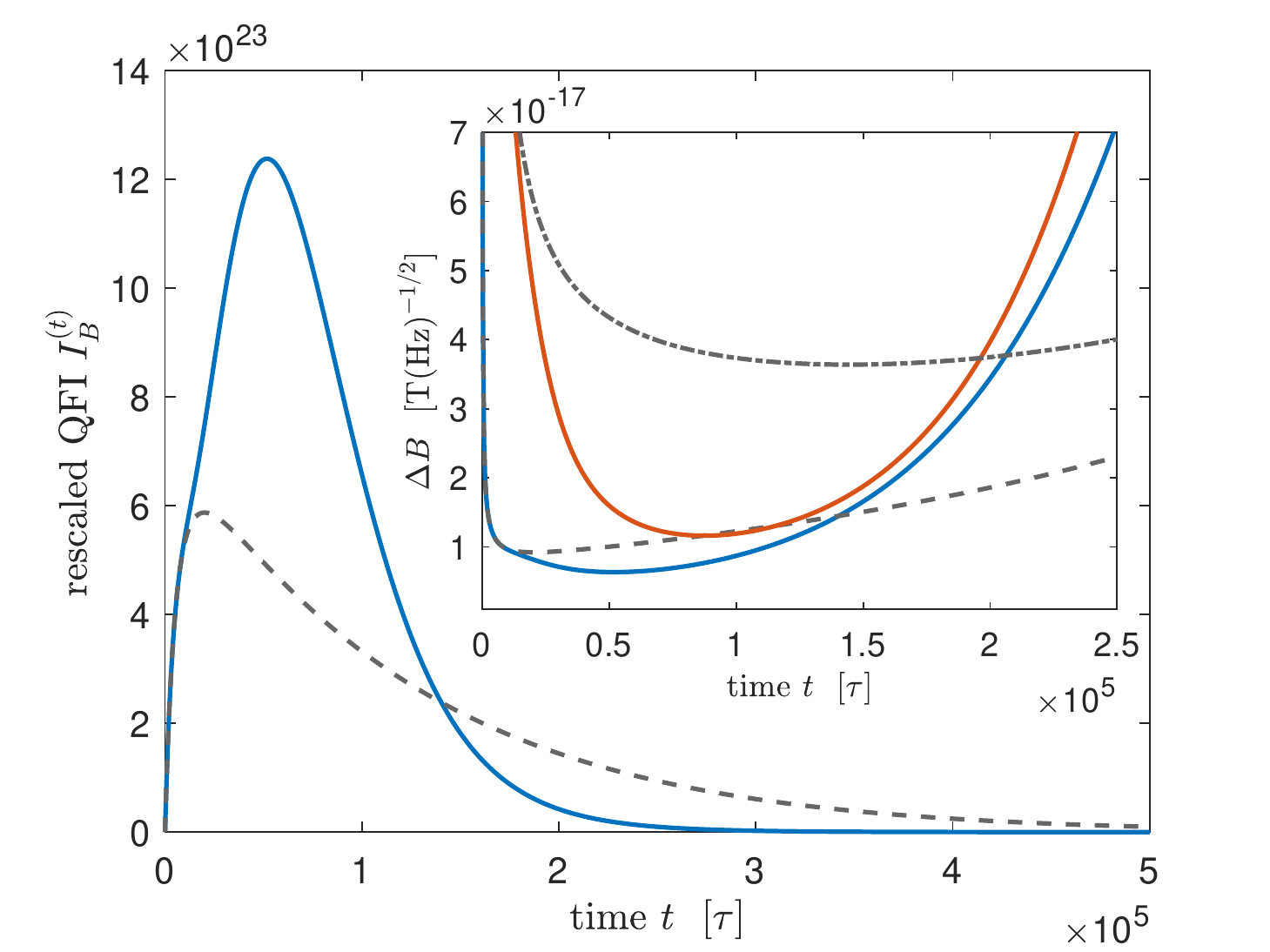}
	\caption{Performance of a magnetic-field measurement with a kicked cesium-vapor magnetometer. The gray dashed line
		shows the rescaled quantum Fisher information $I^{(t)}_B$ for measuring the magnetic field $B$, the blue line is obtained
		by periodically adding short optical kicks every $\tau=1\,$ms. The inset shows
		precision $\Delta B$ in units of T$/\sqrt{\text{Hz}}$ for an
		optimal measurement (gray dashed line and blue line without and with kicks, respectively) and for measuring the $z$-component of the electronic spin $S_z$ (gray dash-dotted and red line without and with kicks, respectively).}\label{fig:SERF}
\end{figure}
We compare now the kicked cesium-vapor magnetometer in the SERF regime as described above (section \ref{sec:master}) with exactly the same magnetometer in the absence of kicks, denoted as reference in the following. Figure \ref{fig:SERF} shows rescaled QFI for both measurement strategies: It is remarkable that the rescaled QFI for the kicked magnetometer continues to increase when the reference already starts to decay due to decoherence although kicks introduce additional decoherence.
The inset shows the measurement precision $\Delta B$ in units of 
T$/\sqrt{\textrm{Hz}}$ per $1\,\textrm{cm}^3$ vapor volume. It is defined by $\Delta B=1/\sqrt{nI_B^{(t)}}$, where $n\simeq 2\times 10^{10}$ is the number
of cesium atoms in $1\,\textrm{cm}^3$, and for a specific measurement
$I_B^{(t)}$ must be replaced by the corresponding rescaled Fisher
information $I_\textrm{Fisher, B}^{(t)}$. We find about $31\%$ improvement 
in measurement precision $\Delta B$ for an optimal measurement and $68\%$ improvement in a comparison of measurements of 
the electronic spin component $S_z$. Such an $S_z$ measurement is easily realized by measuring the  Faraday rotation of an off-resonant readout beam with a polarization measurement as indicated in figure \ref{fig:term} (b).   
\section{DISCUSSION}
We propose a method to improve sensitivity of a cesium-vapor magnetometer in the spin-exchange relaxation-free regime. We do not need to prepare an initial entangled states but supplement the spin-precession dynamics of the magnetometer with nonlinear kicks generated with off-resonant light pulses via the ac Stark effect. The interplay of precession in the magnetic field and non-linear kicks put the dynamics in the regime of quantum chaos. A simulation of the magnetometer based on a master equation including all relevant decoherence mechanisms reveals improved robustness with respect to decoherence by including non-linear kicks to the dynamics even though kicks induce additional decoherence\cite{fiderer2018quantum}. In our concrete example we find up to $68\%$ improvement in measurement precision over a the same magnetometer in the absence of kicks. We expect a further improvement in precision if the transmission of the off-resonant kick pulses is measured. Large improvements can be expected by increasing the system size, which is restricted to single cesium spins in our case because the non-linear kicks act only on single atom spins. Applying the non-linear kicks to the joint spin of cesium atoms in the vapor cell could be realized by exploiting effective interactions between the atoms in a 
cavity as suggested in \cite{agarwal1997atomic} or by using 
the interaction with a propagating light field as demonstrated
experimentally in \cite{takeuchi2005spin,Julsgaard01} with about
$10^{12}$ cesium 
atoms.
Finally, the idea of a quantum-chaotic sensor\cite{fiderer2018quantum} is very general and we expect improvements in many other quantum sensors that can be rendered chaotic as well; candidates are BECs or arrays of NV centers exploiting interactions leading to an effective non-linearity.  
\subsection*{Acknowledgments}
This work was supported by the Deutsche Forschungsgemeinschaft (DFG), Grant No. BR 5221/1-1. Numerical calculations were performed in part with resources supported by the Zentrum f{\"u}r Datenverarbeitung of the University of T{\"u}bingen.
\subsection*{Competing Interests}
The authors declare no competing interests.

\end{document}